\documentclass[12pt,letterpaper]{article}
\usepackage{amsfonts,latexsym,amssymb,amsmath,color,algorithmic,epsfig,rotating}
\newtheorem{dfn}{Definition}[section]

\newtheorem{thm}[dfn]{Theorem} 
\newtheorem{lemma}[dfn]{Lemma}

\newtheorem{ex}[dfn]{Example}

\newtheorem{nota}{Notation}

\newlength{\poonen}
\newcommand{\thth}{^{\text{\underline{th}}}}

\newcommand{\eps}{\varepsilon}

\newcommand{\R}{\mathbb{R}}

\newcommand{\qed}{$\blacksquare$}
\newcommand{\dia}{$\diamond$}

\begin{document}
\title{\mbox{}\\
\vspace{-1in}Sturm's Theorem with Endpoints}  

\author{Philippe P\'ebay$^1$ \and J.\ Maurice Rojas$^2$ \and 
David C.\ Thompson$^1$ } 

\date{\today} 

\maketitle
\footnotetext[1]{ Sandia National Laboratories, e-mail: 
{\tt pppebay@sandia.gov} , {\tt dcthomp@sandia.gov} . } 
\footnotetext[2]{ Department of Mathematics,  
Texas A\&M University
TAMU 3368, 
College Station, Texas 77843-3368,  
USA. 
e-mail: {\tt 
rojas@math.tamu.edu} \ ,    
web page: {\tt 
www.math.tamu.edu/\~{}rojas} \ .
Partially supported by NSF CAREER grant DMS-0349309, and Sandia National 
Laboratories. }  

\begin{abstract}
Sturm's Theorem is a fundamental 19$\thth$ century result relating the number 
of real roots of a polynomial $f$ in an interval to the number of 
sign alternations in a sequence of polynomial division-like calculations. 
We provide a short direct proof of Sturm's Theorem, including the numerically 
vexing case (ignored in many published accounts) where an interval 
endpoint is a root of $f$. 
\end{abstract} 

\section{Introduction} 
Counting the number of roots of a polynomial in an interval is 
a fundamental algorithmic problem in real algebraic geometry, 
and forms the core of techniques for deeper problems such as 
real-solving and the first order theory of the reals. On the practical side, 
numerous problems from control theory and physical modeling 
reduce to solving systems of polynomial equations over the real numbers, 
and one can not solve a system numerically until one understands 
how to count the number of roots in an interval. 

We will present, from scratch, a strengthened 
version of a classical result of Sturm. To begin, let us review 
some basic ideas. 
\begin{nota}
Recall that $\R[x_1]$ is the collection of all polynomials in the 
variable $x_1$ with real coefficients. For any $f\!\in\!\R[x_1]$ of 
degree $d$ we then define its {\em pseudo-remainder sequence} (a.k.a.\ 
{\em Sturm sequence}) to be $P_f\!:=\!(p_0,\ldots,p_d)$, 
where $p_0\!:=\!f$, $p_1\!:=\!f'$ (the derivative 
of $f$),\\
\mbox{}\hfill $p_i\!:=\!q_{i+1}p_{i+1}-p_{i+2}$ 
for all $i\!\in\!\{0,\ldots,d-2\}$,\hfill\mbox{}\\ and $q_{i+1}$ and $-p_{i+2}$ 
are respectively the quotient and remainder obtained from dividing 
$p_i$ by $p_{i+1}$. We also define $P_f(c)\!:=\!(p_0(c),\ldots,p_d(c))$ 
for any $c\!\in\!\R$ and $V_f(c)$ to be the number 
of {\em sign alternations} in the sequence $P_f(c)$. In particular,  
the number of sign alternations in an arbitrary sequence $(s_0,\ldots,s_d)$ 
is simply the number of $j\!\in\!\{0,\ldots,d-1\}$ such that there is a 
$k\!>\!0$ with $s_js_{j+k}\!<\!0$ and $s_\ell\!=\!0$  
for all $\ell$ with $j\!<\!\ell\!<\!j+k$. Finally, 
we let $\sigma : \R \longrightarrow \{-1,0,1\}$ be the {\em sign function},   
which maps all positive (resp.\ negative) numbers to $1$ (resp.\ $-1$) 
and $0$ to $0$. We also naturally extend $\sigma$ to sequences by 
$\sigma(s)\!:=\!(\sigma(s_0),\ldots,\sigma(s_d))$. \dia 
\end{nota} 
\begin{ex}
\label{ex:bad} 
For $f(x_1)\!:=\!x^4_1-2x^2_1+1$, we clearly obtain \\
\mbox{}\hfill $P_f(x_1)\!=\!(x^4_1-2x^2_1+1,4x^3_1-4x_1,x^2_1-1,0,0)$, 
\hfill\mbox{}\\
$\sigma(P_f(-2))\!=\!(1,-1,1)$, $\sigma(P_f(0))\!=\!(1,0,-1)$, and 
thus $V_f(-2)\!=\!2$ and $V_f(0)\!=\!1$. Note also that 
$-1$ is the only root of $f$ in the half-open interval $(-2,1]$, and that 
Sturm sequences can terminate with more than one zero term. \dia 
\end{ex} 
Recall that a root $\zeta$ of $f$ is a {\em multiple} (or {\em degenerate}) 
root iff $f(\zeta)\!=\!f'(\zeta)\!=\!0$. Roots $\zeta$ with 
$f'(\zeta)\!\neq\!0$ are usually called {\em simple} or {\em non-degenerate}. 
\begin{thm}[Refined Sturm's Theorem] 
\label{thm:sturm} 
For any $f\!\in\!\R[x_1]\!\setminus\!\{0\}$ and any real numbers $a$ and $b$ with 
$a\!\leq\!b$, let $N_f(a,b]$ denote the number\footnote{This theorem 
counts {\em distinct} roots and thus does {\em not} count multiple 
roots more than once. }  of roots of 
$f$ in the half-open interval $(a,b]$.\footnote{When 
$a\!<\!b$ this half open 
interval includes $b$ but does {\em not} include $a$, and we use the 
convention $(a,a]\!=\!\emptyset$. European 
authors frequently use $]a,b]$ for what we call $(a,b]$. } 
Then
\begin{enumerate}
\item{If neither $a$ nor $b$ are multiple roots, 
then $N_f(a,b]\!=\!V_f(a)-V_f(b)$.} 
\item{If $f(c)\!=\!f'(c)\!=\!0$ then $V_f(c)\!=\!0$.}
\end{enumerate} 
\end{thm} 

\noindent
Note in particular that we can count roots in $(a,b]$ even when one or both 
endpoints are roots of $f$ --- provided no {\em multiple} roots occur at either 
endpoint. Assertion (2), while almost trivial to prove, is the main reason 
one needs to avoid multiple roots at end-points when using Sturm's Theorem: 
The information carried by the sign alternations of $P_f(c)$ is lost  
entirely when $c$ is a multiple root. 
\begin{ex}
It is easily checked that the roots of $f(x_1)\!:=\!x^3_1-5x^2_1+7x_1-3$
are precisely $\{2,3\}$, with $2$ a multiple root and $1$ a simple root, and that\\
\mbox{}\hfill
$P_f(x_1)\!=\!\left(x^3_1-5x^2_1+7x_1-3,3x^2_1-10x_1+7,\frac{8}{9}x_1-\frac{8}{9},
0\right)$.
\hfill \mbox{}\\
Clearly then, $V_f(1)\!=\!0$, $V_f(2)\!=\!1$, and $V_f(3)\!=\!0$,
so Sturm's Theorem is confirmed for the interval $(2,3]$. However,
we also see that Sturm's Theorem can {\em not} be applied to the
interval $(1,3]$ since $V_f(1)-V_f(3)\!=\!0$ and $f$ in fact
still has a root in $(1,3]$.  \dia
\end{ex}

Curiously, most published accounts 
of Sturm's Theorem avoid considering the presence of any kind of root 
at an endpoint. Furthermore, many accounts assume that $f$ has only 
simple roots. In practice, such an assumption 
can only be enforced by computing square-free parts --- 
a potentially wasteful (and numerically unstable) computation, 
especially when speed is critical. 

The proof of Sturm's Theorem is elementary, but 
is frequently derived as a consequence of more intricate constructions. 
Considering its deep importance in numerical software, we present a 
short direct proof. The theorem follows easily from a single lemma. 
\begin{lemma} 
\label{lemma:key} 
Suppose $f\!\in\!\R[x_1]$ has positive degree $d$, and $c\!\in\!\R$. Then the 
following properties 
hold for the Sturm sequence $P_f(c)\!=\!(p_0(c),\ldots,p_d(c))$: 
\begin{enumerate} 
\item{(Sign Alternation Over a Simple Root) Suppose $p_0(c)\!=\!0$ and 
$p_1(c)\!\neq\!0$. Then for all $\eps\!>\!0$ sufficiently small, 
$\sigma(p_{0}(c-\eps))\!=\!-\sigma(p_{1}(c-\eps))$ 
and $\sigma(p_{0}(c+\eps))\!=\!\sigma(p_{1}(c+\eps))$. } 
\item{(Stability Under Common Multiples) Suppose $p_0$ and $p_1$ are 
each divisible by $g\!\in\!\R[x_1]$. Then 
$P_f\!=\!\frac{1}{g}P_{f/g}$.}  
\item{(Matching Sign Flips Over a Simple Node) Suppose $p_i(c)\!=\!0$ and 
$p_{i+1}(c)\!\neq\!0$ for some $i\!\geq\!1$. 
Then $\sigma(p_{i-1}(c))\!=\!-\sigma(p_{i+1}(c))$. Furthermore, 
for all $\eps\!>\!0$ sufficiently small,\\ 
$\sigma(p_{i-1}(c-\eps))\!=\!-\sigma(p_{i+1}(c-\eps))$ and 
$\sigma(p_{i-1}(c+\eps))\!=\!-\sigma(p_{i+1}(c+\eps))$. } 
\item{(Propagation of Zeroes) For any $i\!\in\!\{0,\ldots,d-2\}$, 
$p_i(c)\!=\!p_{i+1}(c)\!=\!0$ implies that $p_j(c)\!=\!0$ for all $j\!\geq\!i+2$. } 
\end{enumerate} 
\end{lemma}

We are now ready to prove our refined version of Sturm's Theorem.\\ 
{\bf Proof of Theorem \ref{thm:sturm}:}

\medskip 
\noindent 
{\bf Assertion (2):} Since $p_0(c)\!=\!f(c)$ and $p_1(c)\!=\!f'(c)$, 
the recurrence defining the Sturm sequence immediately implies that 
$p_j(c)\!=\!0$ for all all $j\!\geq\!2$, so we are done. 

\medskip 
\noindent 
{\bf Assertion (1):} If $a\!=\!b$, or $f$ is a nonzero constant, then 
we clearly have $N_f(a,b]\!=\!V_f(a)-V_f(b)\!=\!0$ and Assertion (1) indeed 
holds.  So let us assume $a\!<\!b$ and that $f$ has positive degree $d$. 

Let us now reduce to the special case where the following 
condition holds:\\
($\star$) \hfill $(a,b]$ contains at most $1$ root of $f$, 
with $b$ non-degenerate if it is a root of $f$. \hfill \mbox{}\\
To do so, suppose first that $f$ has roots in $(a,b]$ and that, 
in strictly increasing order, they are exactly 
$\zeta_1,\ldots,\zeta_m$. Letting $(c_1,\ldots,c_m)$ be 
any sequence satisfying\\
\mbox{}\hfill $a\!<\!c_1\!<\!\zeta_1\!<\!c_2\!<\!\zeta_2\!<
\cdots <\!c_m\!<\!\zeta_m\!\leq\!b$,\hfill\mbox{}\\ 
we then see that $V_f(a)-V_f(b)$ is exactly\\ \mbox{}\hfill
$(V_f(a)-V_f(c_1))+(V_f(c_1)-V_f(c_2))
+\cdots + (V_f(c_{m-1})-V_f(c_m))+(V_f(c_m)-V_f(b))$.
\hfill\mbox{}\\ 
Since, by definition, $N_f(a,b]$ is exactly \\
\mbox{}\hspace{1.2cm}$N_f(a,c_1]\hspace{.7cm}+\hspace{.7cm}N_f(c_1,c_2]
\hspace{.6cm}+\hspace{.2cm}\cdots\hspace{.2cm} + \hspace{.4cm} 
N_f(c_{m-1},c_m]\hspace{.5cm}+\hspace{.5cm} N_f(c_m,b]$\\
it then clearly suffices to prove\\ 
$N_f(a,c_1]\!=\!V_f(a)-V_f(c_1)$, $N_f(c_1,c_2]\!=\!V_f(c_1)-V_f(c_2)$, \ldots, 
$N_f(c_{m-1},c_m]\!=\!V_f(c_{m-1})-V_f(c_m)$, and 
$N_f(c_m,b]\!=\!V_f(c_m)-V_f(b)$. In other words, whether 
or not $f$ has roots in $(a,b]$, we can indeed assume Condition 
($\star$). 

Let us now reduce even further to the special case where the following 
slightly stronger condition holds:\\ 
($\star\star$) \hfill $(a,b]$ contains at most $1$ root of $f$, and any  
root of $f$ in $(a,b]$ is simple. 
\hfill \mbox{}\\
To do so, observe that $g\!:=\!\gcd(f,f')\!\in\!\R[x_1]$ and, for 
all $i\!\in\!\{1,\ldots,m\}$, $f/g$ is divisible by $x-\zeta_i$ 
but {\em not} divisible by $(x-\zeta_i)^2$. (The latter fact 
follows easily from the product rule for differentation.) So 
$f$ and $f/g$ have the same real roots, except that all the 
real roots of $f/g$ are simple. By Assertion (2) of Lemma \ref{lemma:key} 
we then obtain  
$\sigma(P_f(c)g(c))\!=\!\sigma(P_{f/g}(c))$ and thus 
$\sigma(P_f(c))\!=\!\sigma(g(c))\sigma(P_{f/g}(c))$ for all real $c$. 
In particular, $V_f(c)\!=\!V_{f/g}(c)$ as long as $c$ is not a multiple root. 
Since we are assuming that neither $a$ nor $b$ are multiple roots, 
we can then clearly assume ($\star\star$). 

We are now nearly done: Thanks to Condition ($\star\star$), 
the case where $(a,b]$ contains a unique root $\zeta$  
follows immediately from Assertion (1) of 
Lemma \ref{lemma:key}, assuming we have proved the case 
where $(a,b]$ contains no roots of $f$. For then 
we obtain\\
\mbox{}\hfill $N_f(a,b]\!=\!N_f(a,\zeta-\eps]+N_f(\zeta-\eps,\zeta+\eps]
+N_f(\zeta+\eps,b]\!=\!0+1+0$,\hfill\mbox{}\\ 
where $\eps\!>\!0$ is sufficiently small. 

The final case where $(a,b]$ has no roots requires only 
one more refinement: In our preceding subdivision used to  
enforce Condition ($\star$), suppose we picked {\em more}  
$c_i$, so that the roots of $p_1,\ldots,p_d$ 
(as well as those of $p_0\!=\!f$) were also interlaced. Via the same trick of 
cancellations in an alternating sum, we can thus reduce even further to the 
special case where $(a,b]$ contains no roots of $f$ and contains at most $1$ root 
of $p_1\cdots p_d$. By Assertion (3) of Lemma \ref{lemma:key}, we are done. 
\qed 

\medskip 
\noindent 
{\bf Proof of Lemma \ref{lemma:key}:} Assertions (2) and (4) follow 
immediately from the recurrence defining the Sturm sequence. Indeed, 
upon noting that every $p_i$ is a polynomial linear combination of 
$p_0$ and $p_1$, it is clear that $g|p_0$ and $g|p_1$ together 
imply that every $p_i$ is divisible by $g$. Assertion (2) then 
follows immediately from the uniqueness of remainders in polynomial 
division. For Assertion (4) one merely proceeds by induction. 

To prove Assertion (1) note that if $f'(c)\!=\!p_1(c)\!>\!0$ 
(resp.\ $f'(c)\!=\!p_1(c)\!<\!0$) then 
$f\!=\!p_0$ is locally increasing (resp.\ decreasing) at $c$. 
Since the sign of $p_1$ is locally constant at $c$, Assertion (1) follows 
immediately. 

Assertion (3) follows easily upon observing that $p_i(c)\!=\!0$ 
implies that $p_{i-1}(c)=-p_{i+1}(c)$, thanks to the recurrence defining 
the Sturm sequence. In particular, one need only observe that 
the signs of $p_{i-1}$ and $p_{i+1}$ are locally constant at $c$, 
so $\sigma(p_{i-1}(c-\eps))\!=\!\sigma(p_{i-1}(c))\!=\!\sigma(p_{i-1}(c+\eps))$
and $\sigma(p_{i+1}(c-\eps))\!=\!\sigma(p_{i+1}(c))\!=\!\sigma(p_{i+1}(c+\eps))$. 
\qed 

\section*{Acknowledgements} 
We are most grateful for Jerry Friesen's constant support of this project. 

\bibliographystyle{amsalpha}

\end{document}